\documentclass{webofc}
\usepackage[varg]{txfonts}   
\usepackage{mathtools}
\usepackage{amsmath}
\usepackage{amsfonts}
\usepackage{amssymb}
\usepackage{relsize}
\usepackage{graphicx}
\graphicspath{{Images/}}
\usepackage{hyperref}
\usepackage{slashed}
\usepackage{color}
\usepackage{ulem}

\newcommand{\mev}{\text{ MeV}}

\begin{document}
\title{Chiral restoration of strange baryons}


\author{Eduardo S.\ Fraga \inst{1} \and
Rodrigo da Mata \inst{1}
\and Savvas Pitsinigkos\inst{2}\fnsep\thanks{Speaker}
\and Andreas Schmitt\inst{2}
}

\institute{Instituto de F\'{\i}sica, Universidade Federal do Rio de Janeiro, 
Caixa Postal 68528, 21941-972, Rio de Janeiro, RJ, Brazil
\and
           Mathematical Sciences and STAG Research Centre, University of Southampton, Southampton SO17 1BJ, United Kingdom
          }

\abstract{%
  We review the results of a phenomenological model for cold and dense nuclear matter exhibiting a chiral phase transition. The idea is to model the quark-hadron phase transition under neutron star conditions within a single model, but without adding quark degrees of freedom by hand. To this end, strangeness is included in the form of hyperonic degrees of freedom, whose light counterparts provide the strangeness in the chirally restored phase. In the future, the model can be used for instance to compute the surface tension at the (first-order) chiral phase transition and to study the possible existence of inhomogeneous phases.
}
\maketitle
\section{Introduction and main results}
At high temperatures or high baryon densities, Quantum Chromodynamics (QCD) is expected to form a quark-gluon plasma, while the low-temperature and low-density regime is inhabited by hadrons due to confinement. The intermediate transition regions pose a great theoretical challenge 
due to the difficulty of first-principle calculations, which are currently limited to the regime of very small baryon densities. However, it is this intermediate region that holds the details of quark confinement and the quark-hadron phase transition, along with a microscopic description for the physics of compact stars.

\begin{figure}
\begin{center}
\includegraphics[width=0.9\linewidth]{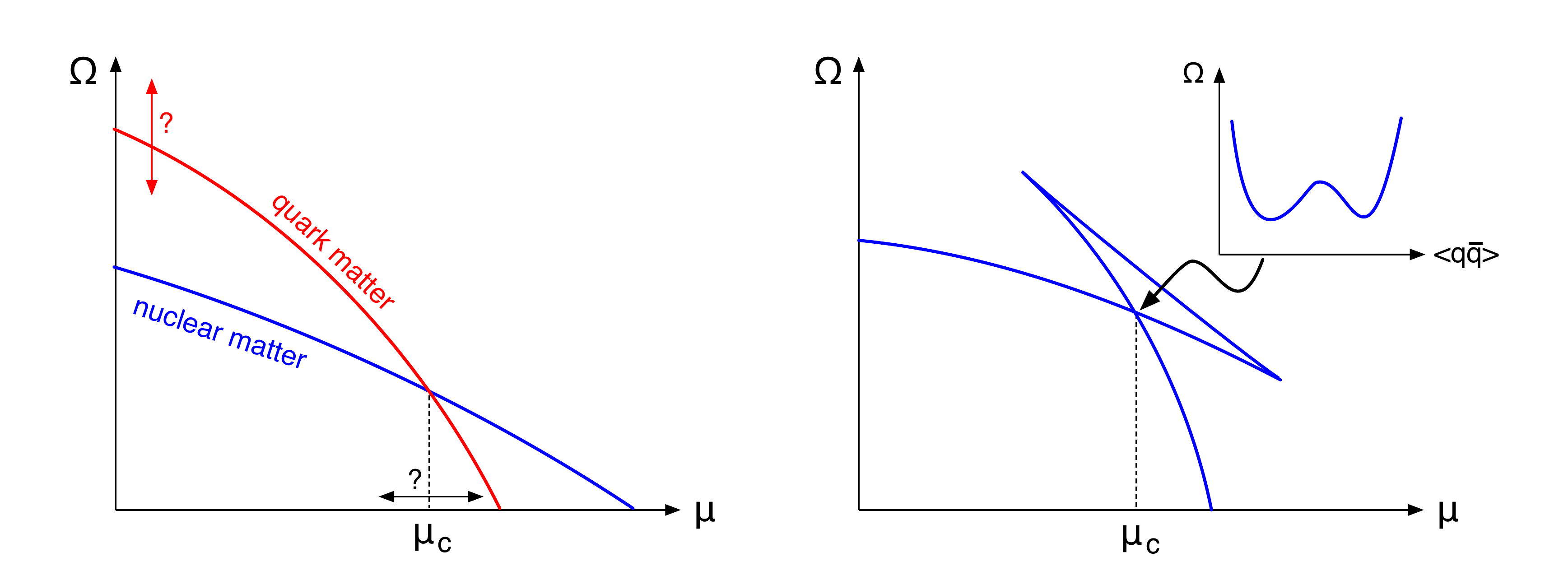}
\vspace{-0.8cm}
\end{center}
\caption{Schematic representation of a two-model (left) and unified (right) approach, showing the quark-hadron transition in the plot of the free energy $\Omega$ vs.\ the chemical potential $\mu$. Using two models, the critical  chemical potential varies as the vacuum pressure of either of the models is changed and thus it is a parameter of the approach, not a prediction. Moreover, a unified approach naturally provides an effective potential connecting the stationary points (inset), which is relevant for instance to compute the surface tension. And, in principle, the single-model allows for a smooth crossover if the two phases are not distinguished by an exact symmetry. }
\label{fig:sketch}
\end{figure}

We explore this interesting region within a phenomenological approach, focusing on zero temperature and large baryon densities. Hadron and quark phases can be described by individual models, but in order to predict the phase transition and its properties we need more than the sum of the two parts. This is why in this work, rather than stitching together two individual models (which can have other benefits \cite{Alford:2013aca,Pereira:2020jgv,Blaschke:2020qqj,Jokela:2020piw,Ferreira:2021osk,Lopes:2021jpm}), we choose a single model approach (as done, in other variants, before  \cite{Marczenko:2020jma,Dexheimer:2020rlp,BitaghsirFadafan:2018uzs,Ishii:2019gta,Kovensky:2020xif}). In particular, we include strangeness in the form of hyperonic degrees of freedom, to account for a more realistic ``quark matter'' phase, extending previous studies, where the only baryonic degrees of freedom in the Lagrangian were neutrons and protons \cite{Drews:2013hha,Drews:2014spa,Fraga:2018cvr,Schmitt:2020tac}. The idea of a unified model for the chiral phase transition is illustrated in Fig.\ \ref{fig:sketch}.

Our model exhibits chiral symmetry restoration at large  baryon densities. This is possible because, by construction, the baryonic mass is entirely generated by the chiral condensate. The chirally restored phase is made of baryonic -- confined -- degrees of freedom, but, due to their lightness, shares some properties with dense quark matter in QCD. 
For instance, we find that we can easily force the system to reproduce the asymptotic limit of the speed of sound, although in our model this limit is approached from above. This is different from perturbative QCD, which is the correct description at asymptotically large baryon densities. We also find that our model parameters can be chosen such that asymptotically dense matter has nonzero strangeness. However, we have not found parameters which reproduce flavor symmetry asymptotically, i.e., a strangeness fraction of 1/3, and at the same time respect astrophysical constraints. Since we include a small explicit chiral symmetry breaking term, our model allows for a chiral crossover, but, again, the constraint of producing realistic masses of compact stars excludes the corresponding parameter regime. We also find that realistic parameter sets do not allow for hyperons to appear in the 
chirally broken phase, i.e., strangeness is only found beyond the chiral phase transition. Finally, we show that putting together all constraints, the model can be used to predict poorly known properties of low-density nuclear matter, such as the slope of the symmetry energy at saturation. These results, and this entire contribution to the proceedings, is based on our previous work \cite{Fraga:2022}.

\section{Lagrangian and approximations}
\label{sec:Lag}

We begin by writing down the Lagrangian of the model. The baryonic part (containing the full baryon octet,  $i=n,p,\Sigma^0,\Sigma^-,\Sigma^+,\Lambda,\Xi^0,\Xi^-$) is 
\begin{equation}
{\cal L}_B = \sum_i\bar{\psi}_i(i\gamma^\mu\partial_\mu+
\gamma^0\mu_i)\psi_i	\, ,		
\end{equation}
where the chemical potentials of the baryons are denoted by $\mu_i$. Three-flavor QCD allows for three independent chemical potentials, and in our main applications this is reduced to one by the constraints of electric neutrality and electroweak equilibrium. 
The mesonic part is
\begin{eqnarray} \allowdisplaybreaks
{\cal L}_M &=& \,\frac{1}{2}\partial_\mu\sigma\partial^\mu\sigma
 - U(\sigma) -\frac{1}{4}\omega_{\mu\nu}\omega^{\mu\nu}-\frac{1}{4}
 \phi_{\mu\nu}\phi^{\mu\nu}-\frac{1}{4}\rho_{\mu\nu}^0
 \rho^{\mu\nu}_0+\frac{m_\omega^2}{2}\omega_\mu\omega^\mu+
 \frac{m_\phi^2}{2}\phi_\mu\phi^\mu\notag\\
 &&+\frac{m_\rho^2}{2}
 \rho_\mu^0\rho^\mu_0
 +\frac{d}{4}(\omega_\mu\omega^\mu+\rho_\mu^0\rho^\mu_0+
 \phi_\mu\phi^\mu)^2		\, ,	 
 \label{eq:Lmeson}			 
\end{eqnarray}
where we have only written the mesons that will form a condensate, where $d>0$ is the quartic self-coupling of the vector mesons, and where the vacuum potential for the chiral condensate is 
\begin{equation}
U(\sigma) = \sum_{n=1}^4 \frac{a_n}{n!} 
\frac{(\sigma^2-f_\pi^2)^n}{2^n}-\epsilon(\sigma-f_\pi) \, ,
\end{equation}
with parameters $a_1, a_2,a_3, a_4, \epsilon$, and the pion decay constant $f_\pi$. The baryon-meson interactions are given by 
\begin{equation}
{\cal L}_I = -\sum_i \bar{\psi}_i(g_{i\sigma} \sigma + g_{i\omega}
\gamma^\mu \omega_\mu +g_{i\rho}\gamma^\mu\rho^0_\mu +g_{i\phi}
\gamma^\mu\phi_\mu )\psi_i		\, , 
\end{equation}
with Yukawa coupling constants $g_{i\sigma}, g_{i\omega},g_{i\rho},g_{i\phi}$. We add non-interacting electrons and muons to the system, characterized by the electron chemical potential $\mu_e$, such that the total Lagrangian is ${\cal L} = {\cal L}_B +{\cal L}_M + {\cal L}_I + {\cal L_\text{leptons}}$. Importantly, there are no explicit mass terms for the baryons. Their medium-dependent  masses $M_i$ are dynamically generated via interactions with the  condensate of the scalar sigma meson $\bar{\sigma}$, which is interpreted as the (non-strange) chiral condensate,  $M_i=g_{i\sigma} \bar{\sigma}$, with $\bar{\sigma}\sim \langle u\bar{u} + d\bar{d}\rangle$. 

Due to the phenomenological nature of the model it is sensible to explore the parameter space, searching for qualitatively different scenarios within the experimentally given constraints. We fix our parameters in terms of:
\begin{itemize}
\item Vacuum masses of baryons and mesons and the vacuum value of the chiral condensate. 

\item Saturation density $n_0 = 0.153\text{ fm}^{-3}$ and corresponding binding energy $E_B = -16.3$ MeV of isospin-symmetric nuclear matter.

\item Incompressibility $K\simeq (200-300)$ MeV, 
symmetry energy $S\simeq (30.2-33.7)$ MeV, and slope 
of the symmetry energy with respect to the baryon density $L\simeq (40-140)$ MeV  at saturation. We fix $S=32\mev$ and $K=250\mev$, having checked that variations of neither of these quantities within the allowed range change our results dramatically. The slope parameter is not fixed a priori but turns out to be constrained indirectly by all our other conditions, as we shall discuss later, see Fig.\ \ref{fig:const}. 
\item Effective nucleon mass at saturation
$M_0\simeq (0.7-0.8)\,m_N$. Our results are very sensitive with respect to variations of this quantity, and we go beyond the empirical range to explore qualitatively different behaviors, see also Fig.\ \ref{fig:const}.
\item Hyperon potential depths in symmetric nuclear matter at saturation, $U_\Lambda^{(N)}=-30$ MeV, and  $U_{\Sigma,\Xi}^{(N)}\simeq -70 
\text{ MeV to }	+30$ MeV. We use these (poorly known) potentials as a guidance to fix our Yukawa couplings, together with theoretical input from a chiral SU(3) approach, see Ref.\ \cite{Fraga:2022} for more details.  
\end{itemize}

We restrict ourselves to zero temperature 
and employ the mean-field and no-sea approximations. We are only interested in thermodynamic equilibrium, and thus our main calculation concerns the free energy density $\Omega$, from which all other relevant quantities can be computed. 
This is done by solving the stationarity equations for the meson condensates together with the equation of local charge neutrality, all under the condition of beta equilibrium. This results in the condensates and $\mu_e$ as (multi-valued) functions of the neutron chemical potential $\mu_n$. Inserting these functions back into the action yields the free energy density $\Omega(\mu_n)$. With standard thermodynamic relations this yields the equation of state, which can be inserted into the Tolman-Oppenheimer-Volkoff equations to compute mass-radius curves of compact stars.

\section{Results}

\begin{figure}
\begin{minipage}{0.49\linewidth}
\includegraphics[width=\linewidth]{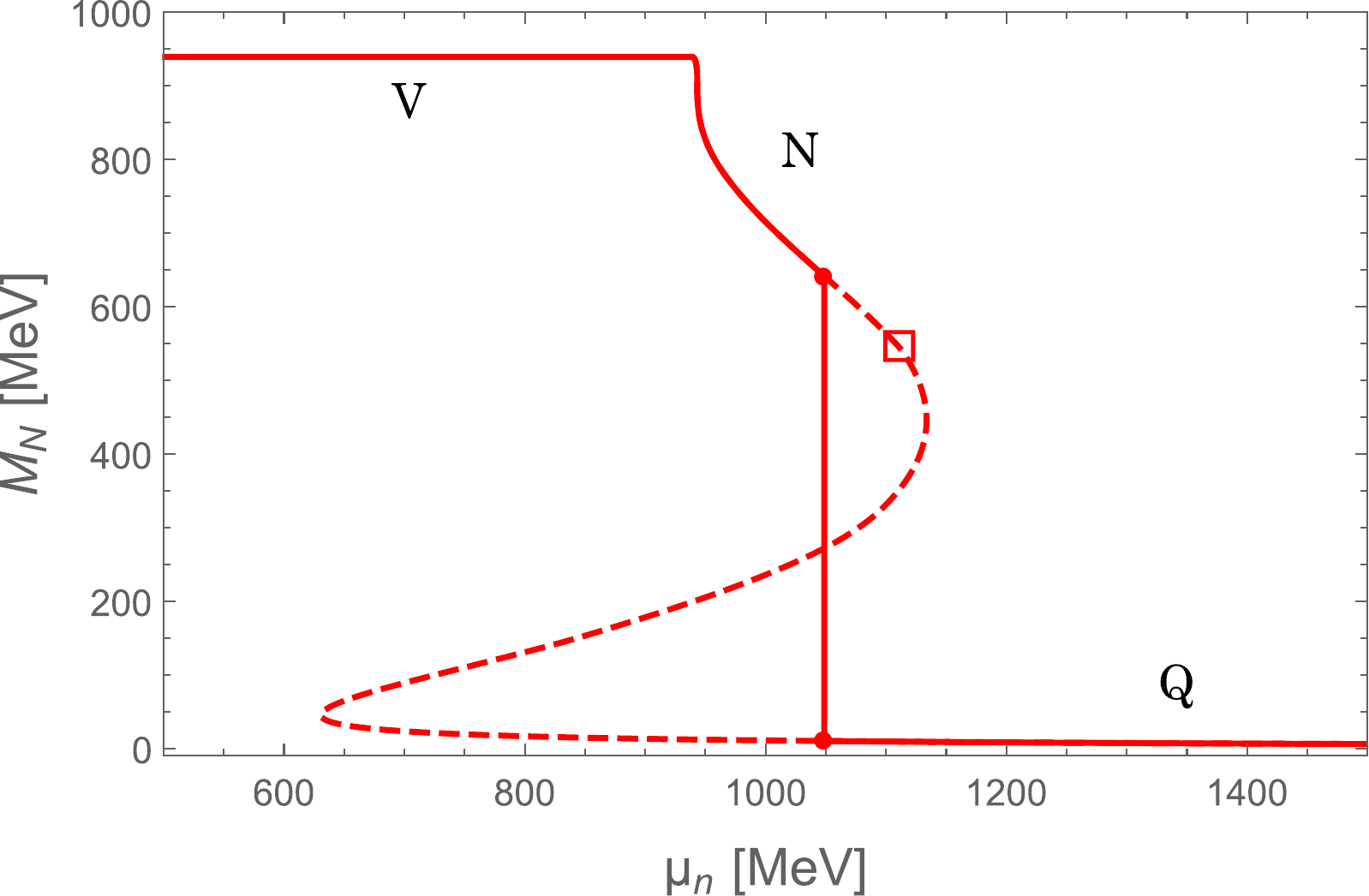}
\end{minipage}
\begin{minipage}{0.5\linewidth}
\includegraphics[width=\linewidth]{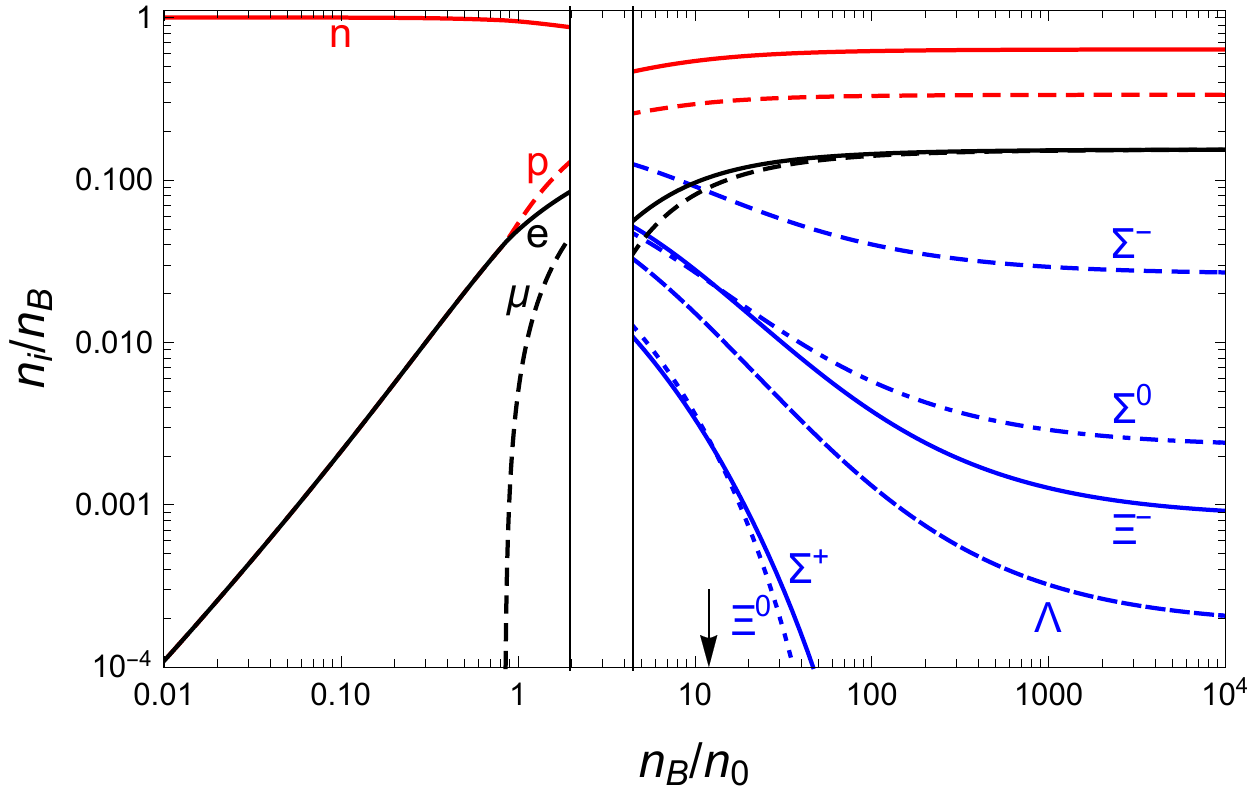}
\end{minipage}
\caption{{\it Left panel:} Effective nucleon mass as a function of the neutron chemical potential. Solid lines represent the stable phases, dashed lines the metastable/unstable phases. The open square marks the appearance of the first hyperonic degree of freedom. Dots mark the chiral phase transition, and the three stable branches are labelled by  vacuum (V), nuclear matter (N) and quark matter (Q). 
{\it Right panel:} Density fractions for each degree of freedom as functions of the total baryon density (normalized by nuclear saturation density). 
The gap in densities reflects the first-order transition.
The arrow marks the density in the core of the heaviest star for this particular parameter choice, which is for both panels $M_0=0.8\, m_N$, $d=21$, $U_{\Sigma,\Xi}^{(N)}=-50\, {\rm MeV}$ and all other parameters as discussed in Sec.\ \ref{sec:Lag}.}
\label{fig:redCase}
\end{figure}

The typical behavior of the system in the presence of a first-order phase transition is best explained with the help of Fig.\ \ref{fig:redCase}. On the left, the effective nucleon mass 
is plotted as a function of the chemical potential. The shape of the curve is valid for all baryons, as their mass is proportional to the nucleon mass.
The plot demonstrates that with increasing chemical potential we go from the vacuum (V) via nuclear matter (N) to a phase with very low nucleon mass, i.e., a phase with approximately restored chiral symmetry, which we interpret as an approximation to quark matter (Q). The first-order nature of the phase transition is manifest in the discontinuity of the effective mass, whose location must be determined from the free energy density. 
The plot also shows the critical chemical potential for the appearance of the first hyperon (open square). For the particular parameters used here, hyperons only appear in the metastable nuclear branch, not in a stable segment of the chirally broken phase. It is only after the chiral symmetry restoration that they can be found in the system, having a very low mass, and thus giving rise to nonzero strangeness of the quark matter branch. This can also be seen in the right panel, which shows that fermions with strangeness (blue) emerge right after the chiral phase transition. 
The particle fractions are shown up to densities much larger than present in neutron stars. This allows us to check whether our model exhibits strangeness asymptotically, as expected from QCD. The curves show   that the asymptotic strangeness fraction is nonzero, but it turns out to be smaller than 1/3, in contradiction to beta-equilibrated, charge neutral QCD, which asymptotes to weakly interacting quark matter with the same number of up, down and strange quarks. We were able to find parameter sets where this flavor symmetry is  recovered, but these do not produce sufficiently massive neutron stars.

\begin{figure}
\begin{minipage}{0.5\linewidth}
\includegraphics[width=\linewidth]{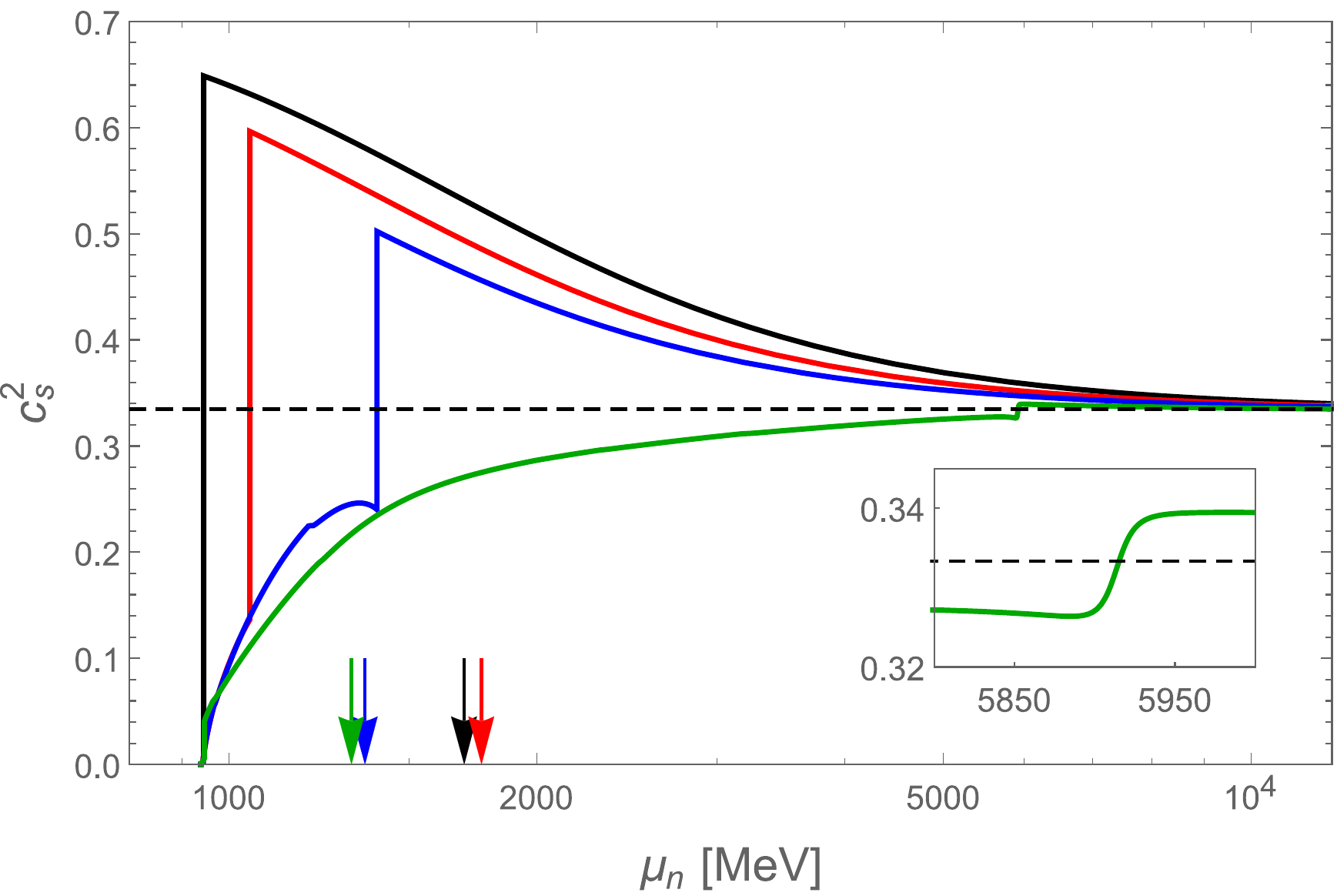}
\end{minipage}
\hspace*{0.04\linewidth}
\begin{minipage}{0.45\linewidth}
\caption{Speed of sound squared as a function of the neutron chemical potential for 4 parameter sets. The  arrows indicate the chemical potential in the center of the heaviest star for each case. The dashed line marks the asymptotic limit $c_s^2 = 1/3$. The red curve corresponds to the parameters of Fig.\ \ref{fig:redCase}, while the others are obtained by varying $M_0/m_N=0.72$ (black), 0.85 (blue), 0.92 (green). }
\label{fig:cs}
\end{minipage}
\end{figure}

In Fig.\ \ref{fig:cs} we plot the speed of sound squared $c_s^2$ for 4 different parameter sets. In each case we see that $c_s^2\to 1/3$  asymptotically, which can be traced back to the presence of the quartic meson coupling $d$  \cite{Fraga:2022}. Our model is of course not asymptotically free and thus interactions still play a role at asymptotic densities. Therefore, the speed of sound does not approach its asymptotic value from below, as predicted by perturbative QCD. 
 The most prominent feature of the speed of sound is the large discontinuity from the chiral phase transition in 3 of the 4 cases, showing a larger speed of sound in the chirally restored phase. Putting together the predictions from perturbative QCD ($c_s^2$ approaching 1/3 from below) and the constraints of heavy neutron stars (requiring $c_s^2>1/3$), it might be more natural to expect a jump {\it down} as we go up in density. However, there is no first-principle argument why this should be the case. It is conceivable that in a more realistic approach than ours, and in QCD itself, $c_s^2$ is non-monotonic in the quark matter phase, such that it can approach 1/3 from below even after having jumped {\it up} at the chiral transition. 
The plot includes a parameter choice that leads to a chiral crossover. As a remnant of the phase transition, the speed of sound still changes rapidly, yet continuously, as shown in the inset. However, as the smallness of the speed of sound already suggests, this case does not allow for  
the observed large masses of neutron stars.

To study the parameter space systematically we enforce the following requirements:
$(i)$ The model should predict compact stars compatible with the current maximum mass limit set by observations, $M_\text{max} =2.1\, M_\odot $ \cite{Fonseca:2021wxt}.
$(ii)$ The asymptotic flavor content of chirally restored matter must include strangeness, as expected in QCD.
$(iii)$ Nuclear matter must be absolutely stable at zero pressure. It is conceivable that nuclear matter is metastable (``strange quark matter hypothesis''  \cite{PhysRevD.30.272,PhysRevD.4.1601}), but in this case quark matter will be thermodynamically preferred at all pressures and there is no quark-hadron transition to study, which is the main motivation of our work.

\begin{figure}
\begin{minipage}{0.49\linewidth}
\includegraphics[width=\linewidth]{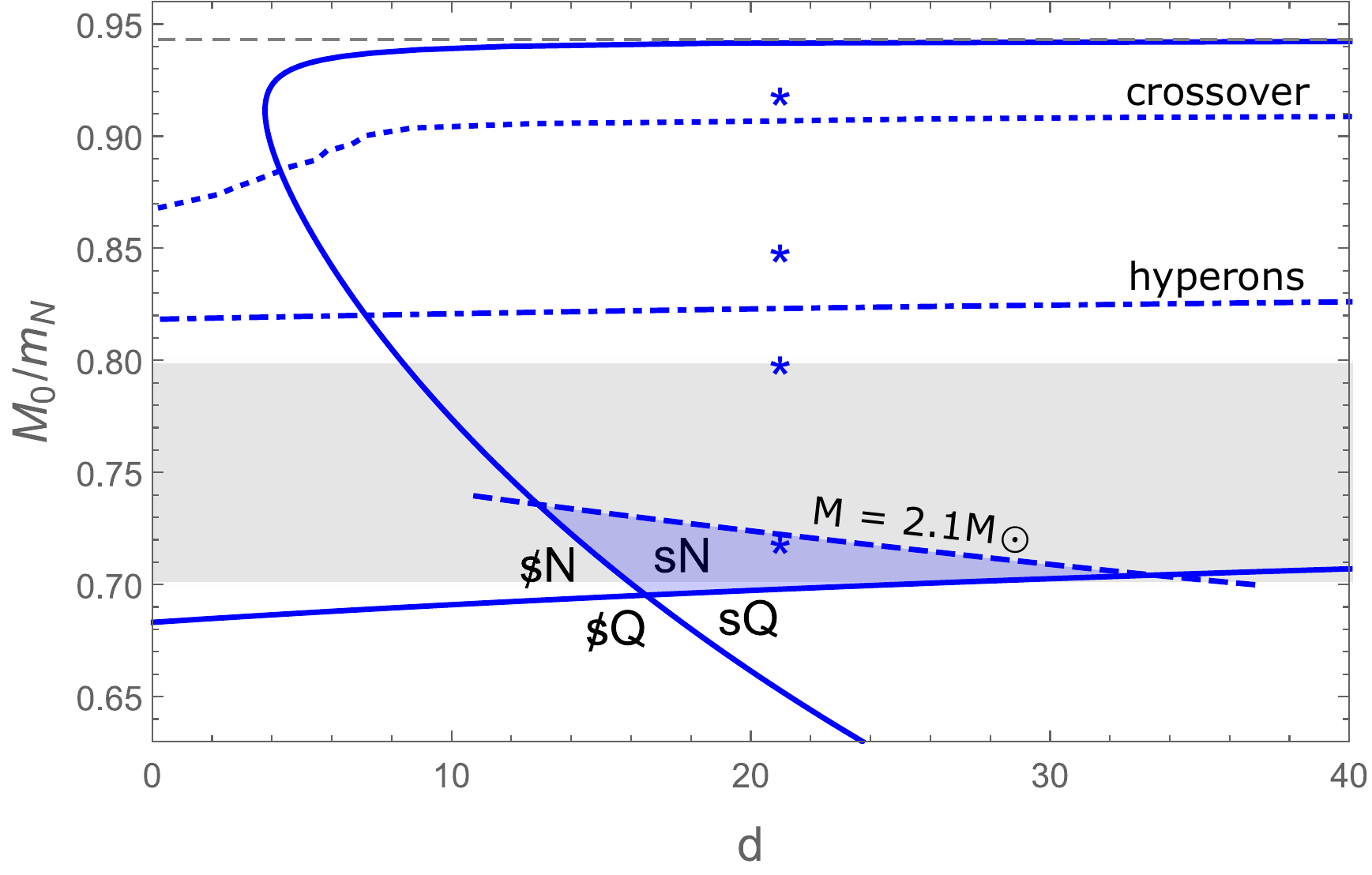}
\end{minipage}
\begin{minipage}{0.49\linewidth}
\includegraphics[width=\linewidth]{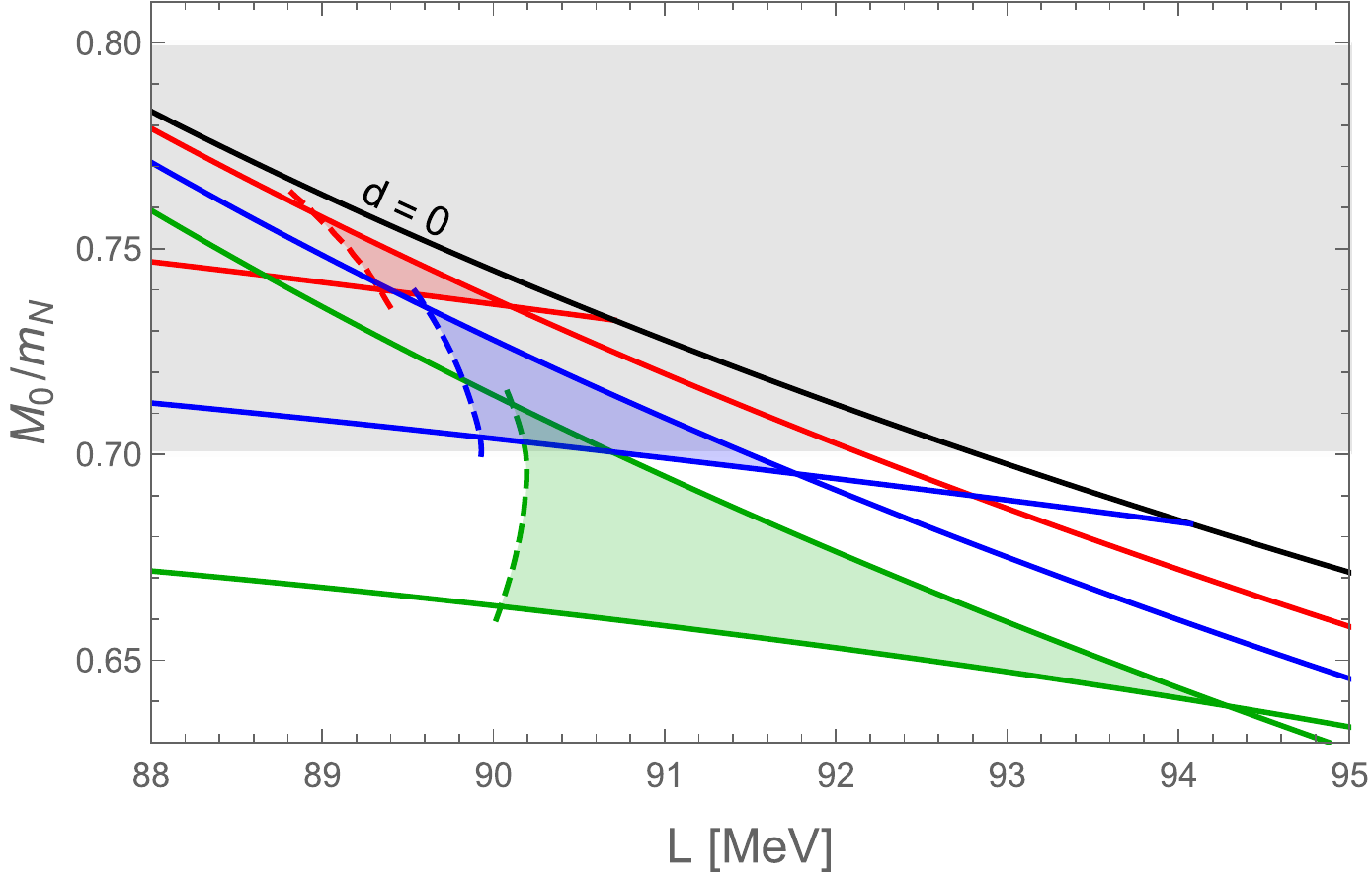}
\end{minipage}
\caption{{\it Left panel}: Allowed region in the $M_0$-$d$ parameter space for $U^{(N)}_{\Sigma ,\Xi}= -50 \mev$. In the 4 areas defined by the solid blue lines, the label (``$\slashed{\text{s}}$'') ``s''  indicates (no) strangeness asymptotically, while N (Q) indicates that  nuclear (quark) matter is stable at zero pressure. The blue dashed curve is the $2.1$ solar mass limit. Above the dash-dotted line hyperons appear before the chiral phase transition, and above the dotted line the quark-hadron transition is a crossover. The gray shaded area is the empirical range for $M_0$. The asterisks mark the 4 choices of Fig.\ \ref{fig:cs}, while the blue shaded area obeys the three criteria listed in the text. {\it Right panel}: The blue curves and area correspond to the left panel, now translated into the $M_0$-$L$ plane. Two additional cases are shown: $U^{(N)}_{\Sigma ,\Xi}= -30 \mev$ (green) and $U^{(N)}_{\Sigma ,\Xi}= -70 \mev$ (red). The $d=0$ line marks the boundary of the parameter space.}
\label{fig:const}
\end{figure}
These 3 criteria can be translated into 3 curves in the $M_0$  - $d$ plane and, equivalently and more physically, the $M_0$ - $L$ plane, see Fig.\ \ref{fig:const}. Together, the curves define an ``allowed'' region, which is shaded in both panels.  Interestingly, the allowed region spans a very narrow range in the slope parameter $L$ range, much narrower than constrained by current experiments, given their uncertainties. We see from the right panel that this region is not significantly enhanced by varying the hyperon potentials $U^{(N)}_{\Sigma ,\Xi}$, and as a result we conclude that our model predicts $L\simeq (88 - 92) \mev$. 
The left panel confirms that the need for heavy stars requires that the chiral phase transition happens before the hyperon onset, and as a result we can only find hyperons in their chirally restored form. This illustrates -- within a single model -- the ``hyperon puzzle'' \cite{Tolos:2020aln} {\it and} its solution: Hyperons should appear at some point as the density is increased but at the same time they seem to prevent stars to become massive. Our model shows that indeed they do appear, but if  strongly coupled quark matter takes over at the chiral phase transition, the hyperon onset can be relegated to a
metastable or even unstable branch, while the stable quark matter core renders the star sufficiently massive. While the quantitative predictions of our phenomenological model should be taken with care, it is remarkable to see the hyperon puzzle and its solution being laid out dynamically within a single model. 

{\it Acknowledgments.} E.S.F.\ is partially supported by CAPES (Finance Code 001), CNPq, FAPERJ, and INCT-FNA (Process No. 464898/2014-5). 
R.M.\ thanks CNPq for financial support.

\bibliography{references}

\begin{thebibliography}{20}

\bibitem{Alford:2013aca}
M.G. Alford, S.~Han, M.~Prakash, Phys. Rev. D \textbf{88}, 083013 (2013)

\bibitem{Pereira:2020jgv}
J.P. Pereira, M.~Bejger, N.~Andersson, F.~Gittins, Astrophys. J. \textbf{895},
  28 (2020)

\bibitem{Blaschke:2020qqj}
D.~Blaschke, A.~Ayriyan, D.E. Alvarez-Castillo, H.~Grigorian, Universe
  \textbf{6}, 81 (2020)

\bibitem{Jokela:2020piw}
N.~Jokela, M.~J\"arvinen, G.~Nijs, J.~Remes, Phys. Rev. D \textbf{103}, 086004
  (2021)

\bibitem{Ferreira:2021osk}
M.~Ferreira, R.~C\^amara~Pereira, C.~Provid\^encia, Phys. Rev. D \textbf{103},
  123020 (2021)

\bibitem{Lopes:2021jpm}
L.L. Lopes, C.~Biesdorf, D.P. Menezes, Mon. Not. Roy. Astron. Soc.
  \textbf{512}, 5110 (2022)

\bibitem{Marczenko:2020jma}
M.~Marczenko, D.~Blaschke, K.~Redlich, C.~Sasaki, Astron. Astrophys.
  \textbf{643}, A82 (2020)

\bibitem{Dexheimer:2020rlp}
V.~Dexheimer, R.O. Gomes, T.~Kl\"ahn, S.~Han, M.~Salinas, Phys. Rev. C
  \textbf{103}, 025808 (2021)

\bibitem{BitaghsirFadafan:2018uzs}
K.~Bitaghsir~Fadafan, F.~Kazemian, A.~Schmitt, JHEP \textbf{03}, 183 (2019)

\bibitem{Ishii:2019gta}
T.~Ishii, M.~J\"arvinen, G.~Nijs, JHEP \textbf{07}, 003 (2019)

\bibitem{Kovensky:2020xif}
N.~Kovensky, A.~Schmitt, JHEP \textbf{09}, 112 (2020)

\bibitem{Drews:2013hha}
M.~Drews, T.~Hell, B.~Klein, W.~Weise, Phys.Rev. \textbf{D88}, 096011 (2013)

\bibitem{Drews:2014spa}
M.~Drews, W.~Weise, Phys. Rev. \textbf{C91}, 035802 (2015)

\bibitem{Fraga:2018cvr}
E.S. Fraga, M.~Hippert, A.~Schmitt, Phys. Rev. D \textbf{99}, 014046 (2019)

\bibitem{Schmitt:2020tac}
A.~Schmitt, Phys. Rev. D \textbf{101}, 074007 (2020)

\bibitem{Fraga:2022}
E.S. Fraga, R.~da~Mata, S.~Pitsinigkos, A.~Schmitt (2022), \texttt{2206.09219}

\bibitem{Fonseca:2021wxt}
E.~Fonseca et~al., Astrophys. J. Lett. \textbf{915}, L12 (2021)

\bibitem{PhysRevD.30.272}
E.~Witten, Phys. Rev. \textbf{D30}, 272 (1984)

\bibitem{PhysRevD.4.1601}
A.R. Bodmer, Phys. Rev. \textbf{D4}, 1601 (1971)

\bibitem{Tolos:2020aln}
L.~Tolos, L.~Fabbietti, Prog. Part. Nucl. Phys. \textbf{112}, 103770 (2020)

\end{thebibliography}

\end{document}